# Elastic Barriers and Formation of Nanoscale Switching Networks

J. C. Phillips

Dept. of Physics and Astronomy, Rutgers University, Piscataway, N. J., 08854-8019

Abstract

Thin films of silicon oxide ($SiO_x$) are mixtures of semiconductive c-Si nanoclusters (NC) embedded in an insulating $g-SiO_2$ matrix. Tour *et al.* have shown that a trenched thin film geometry enables the NC to form semiconductive filamentary arrays when driven by an applied field. The field required to form reversible nanoscale switching networks (NSN) decreases rapidly within a few cycles, or by annealing at 600 C in even fewer cycles, and is stable to 700C. Here we discuss an elastic mechanism that explains why a vertical edge across the planar $Si-SiO_x$ interface is necessary to form NSN. The discussion shows that the formation mechanism is intrinsic and need not occur locally at the edge, but can occur anywhere in the $SiO_x$ film, given the unpinned nanoscale vertical edge geometry.

Many semiconductive materials (such as metal oxides and organic compounds) have exhibited resistive switching, and the phenomenon is appealing for random access memory (RAM) applications. However, the phenomenon has usually been non-reproducibly associated with grain boundaries or electrode interfaces[1]. It has often been assumed that it is exponentially difficult to form native filaments (as distinguished from metallic filaments extended from metallic electrodes) in semiconductive materials. Native filaments can be formed electro-osmotically in Te-based chalcogenide alloys[2] because Te is ambi(2 or 3)-valent, but the switching apparently occurs at electrodes[3]. The presence of metallic filaments formed intrinsically by electron-electron interactions explains[4] the power-law metallic conductivities with α = 0.50 ± 0.01 of insulator-metal transitions in some impurity band semiconductors (Si:P), but these networks are not switchable.

In terms of complementary metal oxide semiconductor (CMOS) technology, by far the most appealing material is $SiO_x$ itself, which is a mixture of semiconductive c-Si 5 nm nanoclusters (NC) embedded in an insulating g-$SiO_2$ matrix. However, it was long supposed that the switching behavior sometimes reported for $SiO_x$ mixtures was associated with conduction through metal filaments from the electrodes[5]. A series of remarkable experiments[5] has shown intrinsic formation of reversible nanoscale switching networks (NSN), provided that an apparently irrelevant geometry is involved: there must be a vertical edge across the planar Si-$SiO_x$ interface. Here we show that this geometrical condition can be explained in terms of long-range elastic forces involving shear induced at the planar Si-$SiO_x$ interface by NSN formation.

Phase separation occurs in the as-deposited planar 40 nm $SiO_x$ (x = 1.9-2) film, but the low-resistive state is formed initially only after an application of V > 20V. For a large enough voltage current paths centered on ~ 5 nm Si nanocrystals (NC) can be activated, but initially these paths will have a complex topology, with knots and the segments generally oriented non-normal to the film. The positions and orientations of the NC are metastabilized by inhomogeneous internal network stresses, which are known to be large in rigid, refractory materials like Si, except for specially selected alloy compositions (such as $[(Na_2O)(SiO_2)_4])$[6]. The knots can be untied, and/or the NC re-oriented, to form more nearly vertical and well-separated filaments in a NSN, or Si can be added by oxygen vacancy diffusion. However, such reconstruction to enhance the electrical conductivity σ increases the internal stress. Because the compressibility is much larger than the shear modulus, NSN formation induces shear, especially at the soft and adaptive planar Si-$SiO_x$ interface. The latter is one of the specially selected alloy compositions which is unstressed at the molecular level, and so can have defect densities of order $10^{-4}$ with a density difference of 30%[7].

The shear stress of the interface provides a large mechanical energy barrier energy for NSN reconstruction, especially when it is associated with a sharp corner at the film edge, where the elastic stress is nearly singular. The significance of corners in place of vertical edges has been demonstrated previously (by shear-induced vertical ordering of ferroelectric nanocrystals in spin-coated thin film polymers[8] and by disappearance of stress singularities at interface edges in vertically nanostructured thin film columnar oxides[9]). It has also been calculated by a finite element analysis for two types of components with different interface edges between the thin film and substrate[9].

In conclusion, the remarkable inducible NSN properties reported for $SiO_x$ mixtures are intrinsic, and are made possible by the absence of singular sheer stress NSN pinning by planar $Si$-$SiO_x$ interfacial corners. Given the rapid evolution of nanoscale science, it may be possible to study these divergent stresses even at the monolayer level, for instance in ceramic high-temperature superconductors[10].

# References

1. Sawa, A. Mater. Today 2008, 11, 28– 36.
2. Ovshinsky, S. R. Phys. Rev. Lett. 1968, 21, 1450-1453.
3. Fritzsche, H. J. Phys. Chem. Sol. 2007, 68, 878-882.
4. Phillips, J. C. Sol. State Comm. 1999, 109, 301-304.
5. Yao, J.; Sun, Z.; Zhong, L.; Natelson, D.; Tour, J. M. Nano Letters 2010, 10, Sept. 3.
6. Kerner, R.; Phillips, J. C. Sol. State Comm. 2001, 117, 47-51 .
7. Lucovsky G.; Wu Y.; Niimi, H.; Misra V.; Phillips, J. C. Appl. Phys. Lett. 1999, 74, 2005-2007.
8. Jung, H. J.; Chang J.; Park, Y. J.; *et al.* Macromol. 2009, 42, 4148-4154.
9. Sumigawa, T.; Hirakata, H.; Takemura, M.; *et al.* Eng. Frac. Mech. 2008, 75, 3073-3083.
10. Phillips, J. C. Proc. Nat. Acad. Sci. (USA) 2010, 107, 1307-1310.